\documentclass[11pt]{article} 
\usepackage{amsfonts,amscd,amsmath}
\def\ep{\text{e}}

\title{Cold Quark Matter, Quadratic Corrections and Gauge/String Duality}
\author{Oleg Andreev\thanks{Also at Landau Institute for Theoretical Physics, Moscow.}
\\ \\
{\it Arnold Sommerfeld Center for Theoretical Physics, LMU-M\"unchen,} \\
{\it Theresienstrasse 37, 80333 M\"unchen, Germany}}
\date{}

\begin{document} 

\vspace{-8cm} 
\maketitle 
\begin{abstract} 
We make an estimate of the quadratic correction in the pressure of cold quark matter using gauge/string duality. 
 \end{abstract}

\vspace{-10cm}
\begin{flushright}
LMU-ASC 01/10
\end{flushright}
\vspace{9cm}

%__________________________  I N T R O ______________________________

\section{Introduction}

The further progress in understanding the phase structure of Quantum Chromodynamics (QCD) at non-zero temperature and density is 
very important in view of the recent experiments at RHIC and also planned at LHC and FAIR. From the theoretical point of view, it can be divided into some regions of "phase space'': perturbative QCD, lattice QCD, phenomenological models, etc. 

As known, perturbative QCD works well at asymptotically high densities \cite{pert-mu} or temperatures \cite{pert-T}, where the QCD coupling $\alpha_s$ is small. For lower densities and temperatures the MIT bag model can be of use \cite{mit}. It is a phenomenological model that effectively includes strong interactions via a bag constant. 

The bag model has taken an interesting turn with the detailed analysis \cite{pisarski} of the lattice data \cite{lattice}. The surprise of this 
analysis is that at zero chemical potential it reveals a term quadratic in temperature as the leading correction to the 
ideal gas term in the pressure. The most recent data \cite{lattice2} indicate that this is generic: that the equation of state of the MIT bag model gets modified as\footnote{Rob Pisarski called it a "fuzzy'' bag model for the pressure \cite{pisarski}.}

\begin{equation}\label{qcd-pressure}
p(T)=aT^4-T^2_\ast T^2-B
\,,\quad\text{with}\quad
T_{\text{\tiny min}}<T<T_{\text{\tiny max}}\,.
\end{equation}
As noted, the novelty is the $T^2$ term, while the remaining two are the standard bag terms, with $B$ a bag constant and $a$ a parameter. A common choice is to take $a$ from perturbation theory up to one loop order. $T_{\text{\tiny min}}$ is close to a critical temperature $T_c$ 
(or some approximate "$T_c$'' for a crossover). A small difference between $T_c$ and $T_{\text{\tiny min}}$ may vary with the 
model. $T_{\text{\tiny max}}$ is set by perturbation theory such that to leading orders it is applicable only for temperatures 
higher than $T_{\text{\tiny max}}$. 

While numerical simulations on the lattice can be of use at non-zero temperature when the quark density is quite small, standard Monte Carlo 
techniques are not of use in cold dense matter because of poor convergence called the sign problem. On the other hand, there are arguments in the literature \cite{viz}
that in QCD long perturbative series (or the UV renormalons) result in the so-called quadratic corrections. From this point of view, the $T^2$ term in \eqref{qcd-pressure} is nothing else but an example of the quadratic correction. If so, then it is natural to expect that at zero temperature the 
equation of state of the MIT bag model can also get modified by a quadratic correction as\footnote{Such a parametrization of the quark matter 
equation of state was first proposed in the context of phenomenology of neutron starts with quark interiors \cite{kapusta} and further discussed in \cite{alford,pisarski-McL}.}

\begin{equation}\label{qcd-pressure2}
p(\mu)=b\mu^4-\mu^2_\ast\mu^2-B
\,,\quad\text{with}
\quad \mu_{\text{\tiny min}}<\mu<\mu_{\text{\tiny max}}\,,
\end{equation}
where again $b$ is a parameter to be fixed from perturbation theory up to one loop order. The $\mu^2$ term is a quadratic correction. 
$\mu_{\text{\tiny min}}$ is expected to be close to a critical value of $\mu$, while $\mu_{\text{\tiny max}}$ is set by leading orders of perturbation theory. Note that $\mu$ stands for the baryon chemical potential, here and below. 

Until recently, the lattice formulation and effective field theories were the main computational tools to deal with nonweakly coupled gauge theories. The situation changed drastically with Maldacena duality (AdS/CFT) \cite{malda} that resumed interest in another tool, string theory. The original duality was for conformal theories, but various perturbations (deformations) produce gauge/string duals with a mass gap, confinement, and chiral symmetry breaking \cite{rev-ads}.	

In this Letter we continue a series of recent studies \cite{a-q2,az13,a-pis} devoted to a search for an effective string theory description of strong interactions. Since precise recipes for finding the string theory dual to QCD are still unknown, our strategy is based on deformations of AdS/CFT. The 
deformation we are pursuing turned out to be successful in providing a systematic approach to the quadratic corrections. Indeed, in \cite{a-q2}, the quadratic correction was found in the two-current correlator. Later, the model was extended to Euclidean signature for computing the heavy 
quark potential, where the quadratic correction occurs as a liner term in the potential at short distances \cite{az13}. Subsequent comparison \cite{cc} with the meson spectrum made it clear that the model should be taken seriously. Moreover, it was also extended to finite temperature. As a result, the $T^2$ term in the pressure \eqref{qcd-pressure} was found \cite{a-pis}. In addition, this model results in the spatial string tension \cite{az1} and the expectation value of the Polyakov loop \cite{az3} which are remarkably consistent with the lattice. Thus, there are reasons to believe that this deformation also provides a good approximation for a string dual to cold quark matter.

%______________________________________________________________________

\section{The Model}

Let us first explain the model to be considered. We take the following ansatz for the 10-dimensional 
background geometry which is a deformed product of the extremal Reissner-Nordstr\"om black hole in Euclidean $\text{AdS}_5$ and a 5-dimensional 
sphere (compact space $X$)\footnote{For completeness, we include a brief summary of the relevant results concerning the Reissner-Nordstr\"om black 
holes in the Appendix.}

\begin{equation}\label{10met}
\begin{split}
ds^2&=\tfrac{l^2}{z^2}H
\bigl(f dt^2+d\vec x^2+f^{-1}dz^2\bigr)+H^{-1}d\Omega_X
\,,\\
H&=\ep^{(1/2)cz^2}\,,\qquad f=\bigl(1-(z/z_+)^2\bigr)^2
\bigl(1+2(z/z_+)^2\bigr)
\,,
\end{split}
\end{equation}
where $z_+=(2/q^2)^{\frac{1}{6}}$. The deformation is due to the same $z$-dependent factor $H$ as those of \cite{a-q2,az13,a-pis}, with $c$ being a parameter whose value will be fixed shortly. Note that \eqref{10met} is smooth and complete at $z=z_+$ such that the inverse period of $t$ is equal to $\beta^{-1}=T=0$, with T the temperature. We also take a constant dilaton and 
discard other (if any) background fields.

Given the background metric \eqref{10met}, we can now find the corresponding gauge potential from the condition of Weyl invariance on a string 
world-sheet. To leading order in $\alpha'$ it is given by 

\begin{equation}\label{A}
\beta_\mu^A=\nabla^\nu F_{\mu\nu}+O(\alpha')=0
\,.
\end{equation}
Here $\beta_\mu^A$ is in fact a renormalization group beta function on the world-sheet. 

For a pure electric potential $A_0(z)$, \eqref{A} becomes

\begin{equation}\label{A1}
\partial_z\Bigl(\sqrt{g}g^{00}g^{zz}\partial_z A_0\Bigr)=0
\,.
\end{equation}
The solution is given by $A_0(z)=C_1\ep^{cz^2}+C_2$, with $C_i$ constants. If we choose the constants so that at $c=0$ 
the solution is reduced to that of Reissner-Nordstr\"om \eqref{RN-A2}, we find 
  
\begin{equation}\label{10A}
	A_0(z)=i\Bigl(-\frac{\sqrt{3}}{2}\frac{q}{c}\bigl(\ep^{cz^2}-1\bigr)+\mu\Bigr)
	\,.
\end{equation}
Finally, we impose the condition $A_0(z_+)=0$ and as a result get $\mu$ as a function of $q$

\begin{equation}\label{barp}
	\mu(q)=\frac{\sqrt{3}}{2}\frac{q}{c}\biggl(\ep^{cz_+^2}-1\biggr)
	\,.
\end{equation}

Following AdS/CFT dictionary \cite{rev-ads}, we identify the parameters $\mu$ and $\rho$, as defined in \eqref{density}, 
with the baryon chemical potential and the baryon number density, respectively.

%_______________________________________________________________________________________________

\section{Estimate of the Quadratic Correction}

Our first goal will be to analyze the non-deformed model. That is, we take $c=0$ for \eqref{10met} and \eqref{10A}. Using the formulas \eqref{10met} and \eqref{density}, we can show that the equation \eqref{barp} yields, up to a constant multiple, a unique solution $\rho(\mu)$. Explicitly, it is given by 

\begin{equation}\label{warmup}
	\rho(\mu)=4b\mu^3
	\,.
\end{equation}
This is the desired result, corresponding in QCD to the fact that for low temperatures and large chemical baryon potentials the baryon number 
density is proportional to the cube of the potential. 

To fix the constant of proportionality, we need some knowledge of the exact string theory dual to QCD or some additional insight. Since the former is beyond our grasp at present, we match the parameter $b$ with that of perturbative QCD neglecting perturbative interactions among the quarks. In doing 
so, we first find the pressure by integrating $\tfrac{dp}{d\mu}=\rho$. In terms of the quark chemical potential $\mu_q=\mu/N_c$, it is $p(\mu_q)=
b N_c^4\mu_q^4$. Finally, we have

\begin{equation}\label{normalization}
	b=\frac{1}{12\pi^2}\frac{N_f}{N_c^3}
	\,.
\end{equation}
Here $N_f$ is the number of quark flavors and $N_c$ is the number of colors. 

Now let us return and discuss the deformed model. At large baryon density (or equivalently at large $q$) it is reasonable to represent \eqref{barp} as 
a series $\tfrac{\sqrt{3}}{2}\tfrac{q}{c}\sum_{n=1}\tfrac{c^n}{n!}\bigl(\tfrac{2}{q^2}\bigr)^{\frac{n}{3}}$. If we take the two leading terms of the series, then 
we can easily invert the function $\mu(q)$. Finally, using the equation \eqref{density}, we find the leading correction to \eqref{warmup}

\begin{equation}\label{rho}
\rho(\mu)=4b\Bigl(\mu^3-\frac{1}{2}\mu_\star^2\mu+O(1)\Bigr)
\,,\qquad
\text{with}
\qquad
\mu_\star=3\sqrt{\frac{c}{2}}
\,.
\end{equation}	

In the homogeneous case the pressure is obtained by integrating the above expression over $\mu$

\begin{equation}\label{Landau}
p(\mu)=b\bigl(\mu^4-\mu_\star^2\mu^2+O(1)\bigr)
\,.
\end{equation}
This is our main result. It includes the $\mu^2$ term, as expected. 

Making an estimate requires some numerics. First, let us consider the light $(u,d)$ quarks. In this case, the value of $c$ is fixed from the 
slope of the Regge trajectory of $\rho(n)$ mesons \cite{regge}. This gives $c\approx0.9\,\text{GeV}^2$ \cite{a-q2}. So, we find for the value of the quadratic correction

\begin{equation}\label{light}
\mu_\star^2\approx 4.1\,\text{GeV}^2
\,.
\end{equation}
In contrast, a simple estimate of the corresponding coefficient of perturbative QCD with $N_f=2$ results in \cite{pert-mu}

\begin{equation}\label{plight}
	\frac{27}{2}(m_u^2+m_d^2)\approx 6\times10^{-4}\,\text{GeV}^2\,.
\end{equation}
Here we have used that $m_u=3\,\text{MeV}$ and $m_d=6\,\text{MeV}$. 

Thus, our model predicts that the $\mu^2$ term being negligible in a pure perturbative region $\mu_{\text{\tiny max}}<\mu$ gets strongly enhanced in the intermediate region $\mu_{\text{\tiny min}}<\mu<\mu_{\text{\tiny max}}$. 

Next, let us discuss the effect of the strange quark. For $N_f=3$, \eqref{plight} becomes $9(m_u^2+m_d^2+m_s^2)$ which is certainly valid near 
the upper limit $\mu_{\text{\tiny max}}$, where $\mu\gg 3m_s$. A simple algebra shows that its value is of order $0.1\,\text{GeV}^2$, with $m_s\approx 0.1\,\text{GeV}$. It is 
still smaller than \eqref{light}, so the effect of the strange quark is not dominant.

Finally, let us estimate the range of $\mu$ for the model of interest. A crude estimate of the lower limit can be made by using the positivity of the 
baryon density and the pressure. It gives that $\mu_{\text{\tiny min}}$ is of order $\mu_\star$. If we assume that like at finite $T$ on the lattice \cite{lattice2}, where $T_{\text{\tiny min}}\sim 1.5T_c$, in the model of interest $\mu_{\text{\tiny min}}\sim 1.5\mu_c$, then using \eqref{light} we arrive at a critical chemical potential of $1.3\,\text{GeV}$ which is reasonable phenomenologically. A crude estimate of the upper limit can be made by 
assuming that at $\mu=\mu_{\text{\tiny max}}$ the contribution of the $\mu^2$ term in the pressure is one order of magnitude smaller than that of
the leading $\mu^4$ term. This gives $\mu_{\text{\tiny max}}\sim 3.3\mu_\star$ or, in terms of $\mu_c$, 
$\mu_{\text{\tiny max}}\sim 5\mu_c$.

%__________________________________________________________________________________________

\section{Concluding Comments}

\hspace{0.5cm} (i) Having derived the equation of state, we can easily develop finite $\mu$ thermodynamics. In particular, for the energy density, we have $\epsilon=b\bigl(3\mu^4-\mu_\star^2\mu^2+O(1)\bigr)$. Combining with \eqref{Landau}, we find the expression for the trace anomaly

\begin{equation}\label{anomaly}
\frac{\epsilon-3p}{\mu^4}=2b\frac{\mu_\star^2}{\mu^2}+O(1)
\,.
\end{equation}
In addition, for the speed of sound $C_s^2=\frac{dp}{d\epsilon}$ we get 

\begin{equation}\label{sound} 
C_s^2(\mu)=\frac{1}{3}\Bigl(1-\frac{1}{3}\frac{\mu_\star^2}{\mu^2}+O(1)\Bigr)
\,.
\end{equation}
All the above formulas are similar to those of \cite{a-pis} at finite $T$.

(ii) Here we used the model based on the deformation of the Reissner-Nordstr\"om solution. Certainly, such a phenomenologically motivated way is out of 
the mainstream of (academic) AdS/CFT, where the background geometry follows from the equations of supergravity and fundamental matter is introduced via 
D-brane embeddings in the probe approximation with $N_c \gg N_f$. One of the advantages of our approach is that it allows us to incorporate the backreaction due to the gauge potential on the background geometry. What really fits better to QCD remains to be seen.

(iii) In the phenomenological parametrization of \cite{alford} the coefficient in front of the $\mu^2$ term arises from the 
strange quark mass as well as color superconductivity. As a result, it is proportional to $m_s^2-4\Delta^2$. Its value is one order of magnitude 
smaller than ours \eqref{light}. 

The formula \eqref{qcd-pressure2} was also suggested, by analogy with the deformed bag model \eqref{qcd-pressure}, 
in the context of Quarkyonic matter \cite{pisarski-McL}. Our interpretation of the $\mu^2$ term as a power correction differs from that of 
\cite{pisarski-McL}, where it is interpreted as due to nonperturbative corrections. However, in the intermediate region of interest some matching conditions between the two regimes may be possible.

\vspace{.25cm}
{\bf Acknowledgments}

\vspace{.25cm}
This work was supported in part by DFG within the Emmy-Noether-Program under Grant No.HA 3448/3-1 and the Alexander von Humboldt Foundation under Grant No.PHYS0167. We would like to thank M. Haack, S. Hofmann and especially V.I. Zakharov for discussions and comments. We also wish to thank J.I. Kapusta for drawing our attention to his work.

\vspace{.35cm} 
%______________________________________________________________________________________________
{\bf Appendix}
\renewcommand{\theequation}{A.\arabic{equation}}
\setcounter{equation}{0}

\vspace{.25cm} 
\noindent In this appendix we review the relevant results concerning the Reissner-Nordstr\"om solutions in five dimensions. 
Most of this material can be found in \cite{RN}.

For the Einstein-Maxwell action with cosmological constant, we take

\begin{equation}\label{action}
	I=-\frac{1}{16\pi G_N}\int d^5x\,\sqrt{g}\biggl(R-l^2F^2+\frac{12}{l^2}\biggr)
	\,.
\end{equation}
Here $G_N$ is the 5-dimensional Newton's constant.

With a pure electric gauge potential\footnote{The parameter $\mu$ is reserved for future use.}

\begin{equation}\label{RN-A}
	A_0(r)=i\Bigl(-\frac{\sqrt{3}}{2l}\frac{q}{r^2}+\mu\Bigr)
	\,,
\end{equation}
a solution of the equations of motion for the metric (with Euclidean signature) takes the spherically symmetric form

\begin{equation}\label{RN-metric}
ds^2=fdt^2+f^{-1}dr^2+r^2d\Omega^2_3
\,,\quad\quad f=1-\frac{m}{r^2}+\frac{q^2}{r^4}+\frac{r^2}{l^2}
\,.
\end{equation}
The parameters $m$ and $q$ are respectively related to the mass and charge of the black hole as

\begin{equation}\label{charges}
	{\cal M}=\frac{3\text{Vol}(\mathbf{S}^3)}{16\pi G_N}m\,,\quad\quad
	Q=\frac{\sqrt{3}\text{Vol}(\mathbf{S}^3)}{4\pi G_N}q
	\,.
\end{equation}
Here $\text{Vol}(\mathbf{S}^3)$ is the volume of a unit 3-sphere.

The solution \eqref{RN-metric} is asymptotic at $r=\infty$ to $\mathbf{S}^3\times\mathbf{S}^1$. A scaling that reduces it to a solution with $\mathbf{R}^3\times\mathbf{S}^1$ may be made as follows. If we introduce a dimensionless parameter $\lambda$ and make the transformation $r\rightarrow\lambda^{\frac{1}{4}}r\,,t\rightarrow\lambda^{-\frac{1}{4}}t\,,m\rightarrow\lambda l^6 m\,,
q\rightarrow\lambda^{\frac{3}{4}}l^5 q\,$, then in the large $\lambda$ limit we obtain 

\begin{equation}\label{RN-metric2}
ds^2=\frac{l^2}{z^2}\left(fdt^2+f^{-1}dz^2+d\vec x^2\right)
\,,\quad\quad f=1-mz^4+q^2z^6\,,
\end{equation}
where $z=l^2/r$. In the process, we have also introduced local coordinates $y_i$ near a point $P\in\mathbf{S}^3$ such that $d\Omega_3^2=\sum dy^2_i$, and then set $x_i=\lambda^{\frac{1}{4}}ly_i$.

Having derived the desired solution for the metric, we can easily obtain that for the gauge potential. From \eqref{RN-A}, we have 

\begin{equation}\label{RN-A2}
A_0(z)=i\Bigl(-\frac{\sqrt{3}}{2}qz^2+\mu\Bigr)
	\,. 
\end{equation}

When we go to $\mathbf{R}^3\times\mathbf{S}^1$, we get that the radius of $\mathbf{S}^3$ is proportional to $\lambda^{\frac{1}{4}}$ 
and so diverges for $\lambda\rightarrow\infty$. Hence, the corresponding volume is also becoming infinite and looks like 
$V_3=\lambda^{\frac{3}{4}}l^3\text{Vol}(\mathbf{S}^3)$. If we introduce the charge density $\rho=Q/V_3$, then the second equation of \eqref{charges} 
becomes

\begin{equation}\label{density}
	\rho=3\sqrt{3}bq
	\,,
\end{equation}
where $b=l^2/(12\pi G_N)$. The difference between $\mathbf{S}^3\times\mathbf{S}^1$ and $\mathbf{R}^3\times\mathbf{S}^1$ is obvious: in the first case $q$ is related to the charge of the black hole, while in the second case it is related to its charge density.  

The metric \eqref{RN-metric2} is smooth and complete if the period of $t$ is $\beta=\frac{4\pi}{\vert f'(z_+)\vert}$, where $z_+$ is the smallest real positive root of $f(z)=0$. 

For $T=0$, the black hole becomes extremal so that $4m^3=27q^4$. In this case the function $f(z)$ takes the form 

\begin{equation}\label{f}
	f=\bigl(1-(z/z_+)^2\bigr)^2\bigl(1+2(z/z_+)^2\bigr)
	\,,\qquad\text{with}\quad z_+=(2/q^2)^{\frac{1}{6}}
	\,.
\end{equation}

%__________________                      R E F S                    ______________________

%____________________________________________________________________
\end{document}